\documentstyle{espart}

\newcommand{\bra}[1]{\left\langle #1 \right|}
\newcommand{\ket}[1]{\left| #1 \right\rangle}

\begin{document}

\begin{frontmatter}

\title{Neutron star properties from modern meson-exchange
potential models}
\author[trond]{G.\ Bao},
\author[oslo]{L.\ Engvik},
\author[oslo]{M.\ Hjorth-Jensen},
\author[oslo]{E.\ Osnes} and
\author[trond]{E. \O stgaard}
\address[trond]{Department of Physics, AVH,
University of Trondheim, N-7055 Dragvoll, Norway}
\address[oslo]{Department of Physics,
University of Oslo, N-0316 Oslo, Norway}

\begin{abstract}
In this work we calculate the total mass, radius, moment of inertia, and
 surface gravitational redshift  for neutron
stars using various
equations of state (EOS). The latter are derived from
the recent meson-exchange potential models of the Bonn group, and we
derive both a non-relativistic and a relativistic EOS.
Of importance here is the fact that relativistic Brueckner-Hartree-Fock
calculations for symmetric nuclear matter meet the empirical data, which
are not reproduced by non-relativistic calculations.
Relativistic
effects are known to be important at high densities, giving an increased
repulsion. This leads  to a stiffer EOS
compared to the EOS derived with a non-relativistic approach.
Both the non-relativistic and the relativistic EOS yield
values for moment of inertia and redshifts in agreement
with the accepted values. The relativistic EOS yields however too
large mass and radius.
The implications are discussed.
\end{abstract}

\end{frontmatter}

\section{Introduction}
The physics of compact objects like neutron stars offers
an intriguing interplay between nuclear processes  and
astrophysical observables \cite{chris92a,wg91}.
Neutron stars exhibit conditions far from those encountered on earth;
typically, expected densities $\rho$ of a neutron star interior are of the
order of $10^3$ or more times the density at neutron ``drip''
$\rho_d\approx 4\cdot 10^{11}$ g/cm$^{3}$.
Thus, the
determination of an equation of state for dense matter is
central to calculations of neutron star properties, and it determines
the mass range as well as the mass-radius relationship for these stars.
It is also an important ingredient in  the determination of the
composition of dense matter  and the thickness of the crust in a
 neutron star.
The latter influences neutrino generating processes and the cooling
of neutron stars \cite{chris92b}. In addition, a theoretical result
for the maximum mass of neutron stars will
have very important astrophysical
implications for the existence and number
of  black holes in the  universe,
 examples are the famous galactic black hole
candidates Cyg X-1 and LMC X-3.
If the maximum mass of a neutron star
 gets smaller, the probability for getting a black hole
after a supernova should become larger.

Important in  this study is the derivation of the equation of state (EOS),
i.e.\ the functional dependence of the  pressure $P$ on density $\rho$,
for dense neutron matter from the underlying many-body theory, derived
from a realistic  nucleon-nucleon (NN) interaction.
By realistic we shall mean a nucleon-nucleon
interaction defined within the framework of meson-exchange theory,
 described conventionally in terms of one-boson-exchange (OBE) models
\cite{mac89,mac93}. Explicitly, we will here build on the
Bonn meson-exchange potential models as they are defined in table
A.2 of ref.\ \cite{mac89}.
Further,
the physically motivated
coupling constants and energy cut-offs which determine the
OBE potentials are constrained through a fit to the
available  scattering data.
The subsequent step is to obtain an effective NN interaction in the nuclear
medium by solving the Bethe-Goldstone equation self-consistently.
Thus, the only parameters
which enter the theory are those which define the NN potential.
Such an approach is commonly referred to as  parameter-free in order
to distinguish it from methods where the meson masses and coupling
constants are adjusted to the bulk nuclear matter properties
\cite{wg91,sw86}.

Until recently, most microscopic calculations of the EOS for
nuclear or neutron matter have been carried out within a
non-relativistic framework \cite{wff88},
where the non-relativistic Schr\"{o}dinger equation is used to describe the
single-particle motion in the nuclear medium. Various degrees of
sophistication are accounted for in the literature \cite{chris92a,mac89},
ranging from first-order calculations in the reaction matrix $G$ to the
inclusion of two- and three-body higher-order
effects \cite{mac89,wff88,dm92,km83}.
A common problem to non-relativistic nuclear matter calculations is,
 however, the
simultaneous reproduction of both the binding energy per nucleon
($BE/A=-16\pm 1$ MeV) and the saturation density with a Fermi momentum
$k_F=1.35\pm 0.05$ fm$^{-1}$.
Results obtained with a variety of methods and nucleon-nucleon (NN)
interactions are located along a band
denoted the ``Coester band'', which
does not satisfy the empirical data for nuclear matter.
Albeit these deficiencies, much progress has been achieved recently
in the description of the saturation properties of nuclear matter.
Of special relevance is the replacement of the non-relativistic
Schr\"{o}dinger equation with the Dirac equation to describe the
single-particle motion, referred to as the Dirac-Brueckner (DB)
 approach.
This is motivated by the success of the
phenomenological Dirac approach in nucleon-nucleus scattering
\cite{ray91} and in the description of properties of finite nuclei
\cite{hnm92}, such as \ the spin-orbit splitting
in finite nuclei \cite{br78}. Moreover, rather promising results
within the framework of the DB approach have been obtained by Machleidt,
Brockmann and M\"{u}ther \cite{bm90,mmb90,lmb92}, employing the
OBE models of the Bonn group. Actually, the empirical properties
of nuclear matter are quantitatively reproduced by Brockmann
and Machleidt \cite{bm90}.

This work falls in four sections. After the
introductory remarks, we
briefly review the general picture in section 2.
In this section we also recast some of the astrophysical equations,
together with the equations of state derived within both the
non-relativistic and the Dirac-Brueckner-Hartree-Fock approaches.
The results are presented  in section 3, while discussions and
conclusions are given in section 4.

\section{General Theory}

In the interior of neutron stars, we find matter at densities
above the neutron "drip" $\rho_d$, and
the properties of  cold dense matter and the associated equation of
state are reasonably well understood at densities up to
$\rho_n\approx 3\cdot 10^{14}$g/cm$^{3}$,
the density at which nuclei begin to dissolve and merge together.
In the high-density range  above $\rho_n$ the physical properties of matter
are still uncertain.

In the region between $\rho_d $ and $\rho_n$ matter
is composed mainly of nuclei,
electrons and free neutrons. The nuclei disappear at the upper end of this
density range because their binding energy decreases with increasing density.
The nuclei then become more neutron-rich and their stability decreases
until a critical value of the  neutron number
is reached, at which point the nuclei dissolve, essentially by merging
together. Since the nuclei present are very neutron-rich, the matter
inside nuclei is very similar to the free neutron gas outside.
However, the external neutron gas reduces the nuclear surface energy
appreciably, and it must vanish when the matter inside
nuclei becomes identical to that outside.

The free neutrons supply an increasingly larger fraction of the total pressure
as the density increases, and at neutron drip the pressure is almost entirely
due to neutrons. Slightly above neutron drip the adiabatic index drops
sharply since the low-density neutron gas contributes appreciably to
the density but not much to the pressure, and it
does not rise again above 4/3 until
$\rho > 7\cdot 10^{12}$ g/cm$^3$. This means that no stable
stars can have central densities in this region.

Relatively "soft" equations of state have been proposed since the
average system energy is attractive at nuclear densities. However,
``stiff'' equations  of state may result from potentials for which the
average system interaction energy is dominated by the attractive
part of the potential at nuclear densities, but by the repulsive part at
higher densities. The stiffer equations of state give rise to important
changes in the structure and masses of heavy neutron stars. As the interaction
energy becomes repulsive above nuclear densities, the corresponding
pressure forces are better able to support stellar matter against
gravitational collapse. The result is that the maximum masses of stars
based on stiff equations of state are greater than those based on soft
 equations  of state. Also,  stellar models based on stiffer equations
have a lower central density, a larger radius and thicker crust.
Such differences are important in determining mass limits for neutron stars,
their surface properties, moment of inertia, etc.

For low densities $\rho < \rho_n$,  where the nuclear force is expected to
be attractive, the pressure is softened somewhat by the inclusion of
interactions. For very high densities, however, the equation of
state is hardened due to the dominance of the repulsive core in
the nuclear potential.

At very high densities above $10^{15}$ g/cm$^3$,
 the composition is expected to include an appreciable number of hyperons
and the nucleon interactions  must be treated relativistically. Relativistic
many-body techniques for strongly interacting matter are, however, not fully
developed.  Presently developed nuclear equations  of state are also
subject to many uncertainties, such as the possibility
of neutron and proton superfluidity, of pion or kaon  condensation, of neutron
solidification, of phase transitions to "quark matter'', and the
 consequences of the $\triangle$ nucleon resonance.

At densities significantly greater
than $\rho_n$, it is no longer
possible to describe  nuclear matter in terms of a non-relativistic
many-body Schr{\"o}dinger equation.
The "meson clouds''  surrounding the
nucleons begin to overlap and the picture of
localized individual particles interacting via two-body
forces becomes invalid. Even before this "break-down" different potentials
which  reproduce reliably low-energy phase shift data result in different
equations of state, since the potentials are sensitive to the repulsive
core region unaffected by the phase-shift data. If the fundamental
building blocks  of all strongly interacting particles are quarks,
any description of nuclear matter at very high densities should involve quarks.
When nuclei begin to ``touch'',  matter just above this
 density should undergo a phase transition at which quarks would
begin to "drip" out
of the nucleons and the result would be quark matter, a degenerate Fermi
liquid.

The main uncertainty  in neutron star models is the equation of state
of nuclear matter, particularly above typical  nuclear densities of $\rho\sim2.
8\cdot10^{14}$
g/cm$^3$. But our present understanding of the condensed matter
is already sufficient  to place quite strict limits  on masses
 and radii of stable neutron stars.

Neutron star models including realistic equations
 of state give the following general result:  Stars calculated with
 a stiff equation of state  have a lower central density, a larger radius,
and a much thicker crust than stars of the same mass computed from
a soft equation of state. Pion or kaon condensation and quark matter would
tend to contract neutron stars of a given mass  and decrease the
maximum mass.

Calculations give the following configurations in the interior:
The surface for $\rho< 10^6$ g/cm$^3$  is a region in which
temperatures and magnetic fields may affect the equation of state.
The outer crust for  $10^6$ g/cm$^3$ $< \rho < 4\cdot 10^{11}$g/cm$^3$
is a solid region where a Coulomb lattice of heavy nuclei coexist in
$\beta$-equilibrium  with a relativistic degenerate electron gas.
The    inner crust for   $4\cdot10^{11}$ g/cm$^3$ $< \rho <
2\cdot10^{14}$g/cm$^3$
consists of a lattice of neutron-rich nuclei together with a superfluid
 neutron gas and an electron gas. The neutron liquid for
 $2\cdot  10^{14}$ g/cm$^3$ $< \rho < 8\cdot10^{14}$g/cm$^3$ contains mainly
superfluid neutrons with a smaller concentration of superfluid
protons and normal electrons. The core region for $\rho> 8\cdot10^{14}$
g/cm$^3$ may not exist in some stars and will depend on the existence
of pion or kaon condensation, neutron solid, quark matter, etc. The
existence of "quark stars'' also remains a possibility.

The minimum mass of a stable neutron star  is determined by
setting the mean value of the adiabatic index  $\Gamma$ equal to the
critical value
for radial stability against collapse. The resulting minimum mass is
$M\sim 0.1M_{\odot}$, where $M_\odot $ is the solar mass, with a
corresponding central density of $\rho\sim
10^{14}$ g/cm$^3$ and radius $R\sim 200 $ km.
The maximum mass equilibrium configuration is somewhat uncertain,
but all microscopic calculations give $M<2.7M_\odot$.

Astronomical observations leading to global neutron star parameters
such as total mass, radius, or moment of inertia,  are important since they
are sensitive to microscopic model calculations.
The most reliable way of determining masses is
 via Kepler's third law in binary pulsars. Observations of such pulsars give
approximately
a common mass region consistent with all data of
$ 1.2 M_\odot < M< 1.8M_\odot$. Present mass determinations for neutron
 stars are  all consistent with present stellar evolution theories.

A general limit for the maximum mass can
 be estimated by assuming the following: General
 relativity is the correct theory of gravitation. The equation of state must
 satisfy the "microscopic stability'' condition $dP/d\rho \geq 0$
and the causality condition $dP/d\rho <c^2$, and should match
some known low-density  equation of state. This gives an upper limit of
$M\sim3-5
M_\odot$. Stiff equations of state in calculations predict
a maximum mass
in the range $M\sim1.5-2.7
M_\odot$. Rotation may
increase the maximum neutron star mass, but not appreciably, i.e., $< 20\%$.

We aim here at discriminating between
equations of state for pure neutron matter
derived from non-relativistic and relativistic
approaches (to be discussed below).
As discussed above, relativistic effects become important at densities
higher than $\rho_n$, and it is therefore of interest to understand
whether the two approaches yield significantly different neutron star
properties. The derivation of the equations of state is discussed
in the first subsection, whereas the equations which define the
calculations of mass, radius, moment of inertia and gravitational
redshifts are discussed in the subsequent subsection.

\subsection{Derivation of the equation of state for neutron matter}

A determination of the equation of state for neutron matter
from the underlying many-body theory has been the subject of
much effort for many years without a general consensus
\cite{glen88,wgw91} on its behaviour at densities higher than
the density of normal nuclear matter.  It is, however, hoped that
bulk properties of neutron stars as those discussed in the previous
section, can shed some light on the functional dependence of the
energy per particle ${\cal E}/A$.

In this subsection we discuss both a non-relativistic and a relativistic
equation of state for neutron matter within the framework of
the Brueckner-Hartree-Fock (BHF) approach. Especially, in this
work we study the BHF approach as it is approximated by the
model-space BHF method of Kuo and Ma \cite{km83}.

\subsubsection{The model-space Brueckner-Hartree-Fock approach}

The basic formalism behind the
model-space BHF approach has been exposed elsewhere, see e.g.,
refs.\ \cite{km83,shk87,jkm89}, thus  we will here only briefly sketch
the essential ingredients which enter our calculations.

Following the conventional many-body approach, we divide the full
hamiltonian $H=T+V$, with $T$ being the kinetic energy
and $V$ the bare NN potential,
into an unperturbed part $H_0 =T+U$ and an interacting part $H_I = V-U$,
such that
\[
   H=T+V=H_0 + H_I,
\]
where we have introduced an auxiliary single-particle (sp)  potential $U$. If
$U$ is chosen such that $H_I$ becomes small, then perturbative
many-body techniques can presumably be applied.
A serious obstacle to any perturbative treatment is the fact that the
bare NN potential $V$ is very large at short inter-nucleonic distances,
which renders a perturbative approach highly prohibitive. To overcome
this problem, we introduce the reaction matrix $G$ given
by the solution of
the Bethe-Goldstone equation (in operator form)
\begin{equation}
   G(E)=V+VQ\frac{1}{E - QH_0Q}QG,
   \label{eq:bg}
\end{equation}
where $E$ is the energy of the interacting  nucleons
and $Q$ is the Pauli
operator which prevents scattering into occupied states.
More explicitly, the above equation reads (in a partial wave
representation)
\begin{equation}
   G_{ll'}^{\alpha}(kk'KE)=V_{ll'}^{\alpha}(kk')
   +\sum_{l''}\int \frac{d^3 q}{(2\pi )^3}V_{ll''}^{\alpha}(kq)
   \frac{Q(q,K)}{E-H_0}
   G_{l''l'}^{\alpha}(qk'KE),
   \label{eq:gnonrel}
\end{equation}
with $ll'$ and $kk'$ the orbital angular momentum and the linear
momentum of the
relative motion, respectively. $K$ is the momentum of the center-of-mass
motion. Since we are going to use an angular
average for the Pauli operator, the $G$-matrix is diagonal in total
angular momentum $J$ and orbital angular momentum $L$ for the center-of-mass
motion. Further, the $G$-matrix is diagonal in isospin $T$ and spin $S$.
These quantities, i.e. $J$, $L$, $T$ and $S$, are all represented
by the variable $\alpha$. The term $H_0$ in the denominator of eq.
(\ref{eq:gnonrel}) is the unperturbed energy of the intermediate states,
and $K$ is the corresponding momentum of
the center-of-mass motion. $H_0$ depends on both $k$ and $K$, see
discussion below.
The $G$-matrix elements are anti-symmetrized.

The choice of the  Pauli operator is decisive to the determination of the
sp
spectrum. Basically, to first order in the reaction matrix $G$
there are three commonly used sp spectra, all
defined by the self-consistent solution of the following equations
\begin{equation}
   \varepsilon_{\alpha} = \varepsilon (k_{\alpha})
   = t_{\alpha} + u_{\alpha}=\frac{k_{\alpha}^2}{2m}
   +u_{\alpha},
   \label{eq:spnrel}
\end{equation}
where $m$ is the bare nucleon mass,
and
\begin{equation}
   \begin{array}{ccc}\\
   u_{\alpha} =& {\displaystyle \sum_{h \leq k_F}}
   \left\langle \alpha h \right|
   G(\omega = \varepsilon_{\alpha} + \varepsilon_h )
   \left| \alpha h \right\rangle _{\mathrm{AS}}, & k
    \leq k_M,  \\ \\
   u_k=&0,& k > k_M. \end{array}
   \label{eq:selfcon}
\end{equation}
For notational economy, we set $|{\bf k}_{\alpha}|=k_{\alpha}$.
Here we use anti-symmetrized matrix elements (AS), and $k_M$ is a cut-off
on the momentum. $t_{\alpha}$ is the sp kinetic energy and similarly
$u_{\alpha}$
is the
sp potential.
The choice of cut-off $k_M$ is actually what determines the three
commonly used sp spectra.
In the conventional BHF approach one takes $k_M = k_F$, which leads
to a Pauli operator $Q_{\mathrm{BHF}}$
(in the laboratory system) given by
\begin{equation}
   Q_{\mathrm{BHF}}(k_m , k_n ) =
    \left\{\begin{array}{cc}1,&min(k_m ,k_n ) > k_F,\\
    0,&otherwise.\end{array}\right.
    \label{eq:bhf}
\end{equation}
The BHF choice sets $u_k = 0$ for $k > k_F$, which leads
to an unphysically large gap at the Fermi surface. To overcome this
problem, Mahaux and collaborators \cite{ms89}
introduced a continuous sp spectrum
for all values of $k$, i.e. they set $k_M = \infty$. The divergencies
which then may occur in eq.\ (\ref{eq:bg}) are taken care of  by
introducing
a principal value integration in eq.\ (\ref{eq:bg}), to retain only the
real part contribution to the $G$-matrix.

Finally,
the model-space BHF approach which we shall employ, adopts a cut-off
 $k_M =  ak_F$,
where $a$ is a constant. Thus, the cut-off $k_M$ is given
as a multiple
of the Fermi momentum. A frequently used value \cite{km83}
is $a= 2$,
a choice also used here. This means that we extend
the BHF spectrum
to go beyond $k_F$. The resulting Pauli operator $Q_{\mathrm{MBHF}}$
( in the laboratory system)
for the model-space BHF method is
\begin{equation}
   Q_{\mathrm{MBHF}}(k_m , k_n ) =
     \left\{\begin{array}{cccc}1,
     &min(k_m ,k_n ) > k_F&and&max(k_m ,k_n ) > k_M, \\
     0,&&otherwise.&\end{array}\right.
     \label{eq:mbhf}
\end{equation}
If we set $k_M = k_F$ , we obtain the traditional BHF choice.
Since we use an angle-average approximation to the Pauli operator, it
is convenient to transform the Pauli operator from the laboratory
frame of reference to that of the center-of-mass and relative motion.
For these details, we refer the reader to refs.\ \cite{km83,shk87}.

In connection with the model-space BHF method,
it is worth noting the following:
Relating the model-space BHF approach to the conventional BHF sp spectrum
and the continuous sp spectrum, we  may say that the model-space BHF is
an intermediate
scheme
in the sense that we introduce an extended Pauli operator in eq.\
(\ref{eq:mbhf}) such that we have a continuous sp spectrum for $k < k_M$,
while for $k>k_M$ the spectrum is that of a free particle. Moreover, the
model-space BHF definition of the Pauli operator gives
a $G$-matrix which does
not account for scattering into intermediate states if both particles
have momenta between $k_F$ and $k_M$. This is, however, a welcome feature
of the model-space BHF
method as it allows one to consider collective excitations
like the summation of particle-particle hole-hole (pp-hh) terms
without double-counting problems \cite{shk87}.
Finally, a more general approach to eq.\ (\ref{eq:selfcon}) would be
to replace the $G$-matrix with the self-energy vertex function,
which
includes higher-order effects in $G$ as well. However,
as already pointed
out above, we will limit our attention to $G$ only.

It should be remarked that although we have removed the angle
dependence of the Pauli operator, the energy
denominator in eq. (\ref{eq:gnonrel})
still depends on the angle between the relative and
center-of-mass momenta. This angle dependence is handled by the
so-called effective mass approximation. The single-particle energies
in nuclear matter are assumed to have the simple quadratic form
\begin{equation}
  \begin{array}{ccc}
   \varepsilon_{\alpha}=&{\displaystyle\frac{k_{\alpha}^2}
   {2m^{*}}}+\Delta,&k_{\alpha}\leq k_M,\\
   &&\\
  =&{\displaystyle\frac{k_{\alpha}^2}{2m}},&k_{\alpha}> k_M ,\\
  \end{array}
  \label{eq:spen}
\end{equation}
where $m^{*}$ is the non-relativistic
effective mass of the nucleon and $m$ is the
bare nucleon mass. For particle states above $k_M$ we choose
a pure kinetic energy term, whereas for hole states
the terms $m^{*}$ and $\Delta$ ($\Delta$ is an effective single-particle
potential related to the $G$-matrix) are obtained through the
self-consistent Brueckner-Hartree-Fock (BHF) procedure. This
self-consistency scheme consists in choosing adequate initial values of the
effective mass and $\Delta$. The obtained $G$-matrix is then used to
obtain new values for $m^{*}$ and $\Delta$, and this procedure
continues until these parameters show little variation.
The starting energy $E$ is then determined by
the energy of the interacting nucleons, i.e.,
\[
E =\frac{k^2}
{m^{*}}+\frac{K^2}
{4m^{*}}+2\Delta.
\]

Finally, the non-relativistic
energy per particle ${\cal E}/A$ is formally given as
\begin{equation}
   {\cal E}/A =
   \frac{1}{A}\left\{ \sum_{h\leq k_F}
   \frac{k_h^2}{2m}+
   \frac{1}{2}\sum_{hh'\leq k_F}
   \bra{hh'}G(E=\varepsilon_h +\varepsilon_{h'})
   \ket{hh'}_{\mathrm{AS}}\right\}.
   \label{eq:enrel}
\end{equation}

\subsubsection{Relativistic effects}

The properties of neutron stars depend on the equation of state
at densities up to an order of magnitude higher than those observed
in ordinary nuclei. At such densities, relativistic effects
certainly prevail. Among relativistic approaches to the nuclear
many-body problem, the so-called Dirac-Hartree and Dirac-Hartree-Fock
approaches have received much interest.
One of the early successes of these
approaches was the quantitative reproduction of spin observables,
which are only poorly described by the non-relativistic theory.
Important to these methods
was the introduction of a strongly attractive scalar component
and a repulsive vector component \cite{sw86,br78}
in the nucleon self-energy. Inspired
by the successes of the Dirac-Hartree-Fock method, a relativistic
extension of Brueckner theory was proposed by the Brooklyn group
\cite{brook}, known as the Dirac-Brueckner theory.
One of the appealing features of the Dirac-Brueckner approach
is the self-consistent determination of the relativistic sp
energies and wave functions.
The Dirac-Brueckner approach differs from the Dirac-Hartree-Fock one
in the sense that in the former one departs from the free NN potential
which is only constrained by a fit to the NN data, whereas the
Dirac-Hartree-Fock method pursues a line where the parameters of the
theory are determined so as to reproduce the bulk properties
of nuclear matter. It ought, however, to be stressed that the Dirac-Brueckner
approach \cite{bm90,brook,hm87},
which starts from NN potentials based on meson exchange,
is a non-renormalizable theory, where the short-range part
of the potential depends on additional parameters like
vertex cut-offs, clearly minimizing the
sensitivity of calculated results to short-distance inputs. This
should be contrasted to the Dirac-Hartree-Fock
method pioneered by Walecka and Serot \cite{sw86,hs87,ser92}.

The description presented here for the Dirac-Brueckner approach follows
closely that of Brockmann and Machleidt \cite{bm90}. We will thus
use the meson-exchange models of the Bonn group, defined
in table A.2 of ref.\ \cite{mac89}. There the three-dimensional
reduction of the Bethe-Salpeter equation as given by the Thompson equation
\cite{thomp70} is used to solve the equation for the scattering
matrix. Hence, including the necessary medium effects like the
Pauli operators discussed in the previous subsection and the starting
energy, we will rewrite eq. (\ref{eq:gnonrel}) departing from
the Thompson equation. Then, in a self-consistent way, we
determine the above-mentioned scalar and vector components  which
define the nucleon self-energy. In this sense we also differ
from the non-relativistic approach discussed above, where the
parameters which are varied at each iterative step are the
non-relativistic effective mass and the effective sp potential
$\Delta$.

In order to introduce the relativistic nomenclature, we consider first the
Dirac equation for a free nucleon, i.e.,
\[
  (i\not \partial -m )\psi (x)=0,
\]
where $m$ is the free nucleon mass and $\psi (x)$ is  the nucleon field
operator
($x$ is a four-point)
which is conventionally expanded in terms of plane wave states
and the Dirac spinors $u(p,s)$, and $ v(p,s)$, where
 $p=(p^0 ,{\bf p})$ is
a four momentum\footnote{Further notation is  as
given  in
Itzykson and Zuber
\cite{iz80}.
Moreover, hereafter we set $G=c=\hbar=1$, where
$G$ is the gravitational constant.} and  $s$ is the spin projection.

The positive energy Dirac spinors are
(with $\overline{u}u=1$)
\begin{equation}
   u(p,s)=\sqrt{\frac{E(p)+m}{2m}}
	  \left(\begin{array}{c} \chi_s\\ \\
	  \frac{\mbox{\boldmath $\sigma$}\cdot{\bf p}}{E(p)+m}\chi_s
	  \end{array}\right),
   \label{eq:freespinor}
\end{equation}
where $\chi_s$ is the Pauli spinor and $E(p) =\sqrt{m^2 +|{\bf p}|^2}$.
To account for medium modifications to the free Dirac equation,
we introduce the notion of the self-energy $\Sigma (p)$.
As we assume parity to be a good quantum number, the self-energy of a
nucleon can be formally written as
\[
       \Sigma(p) =
       \Sigma_S(p) -\gamma_0 \Sigma^0(p)
       +\mbox{\boldmath $\gamma$}{\bf p}\Sigma^V(p).
\]
The momentum dependence of $\Sigma^0$
and $\Sigma_S$ is rather weak \cite{sw86}.
Moreover, $\Sigma^V << 1$, such
that the features of the Dirac-Brueckner-Hartree-Fock
procedure can be discussed within the framework of the phenomenological
Dirac-Hartree ansatz, i.e. we  approximate
\[
\Sigma \approx \Sigma_S -\gamma_0 \Sigma^0 = U_S + U_V,
\]
where $U_S$ is an attractive
scalar field and $U_V$ is the time-like component
of a repulsive vector field.
The finite self-energy modifies the
free Dirac spinors of eq. (\ref{eq:freespinor}) as
\[
   \tilde{u}(p,s)=\sqrt{\frac{\tilde{E}(p)+\tilde{m}}{2\tilde{m}}}
	  \left(\begin{array}{c} \chi_s\\ \\
	  \frac{\mbox{\boldmath $\sigma$}\cdot{\bf p}}
          {\tilde{E}(p)+\tilde{m}}\chi_s
	  \end{array}\right),
\]
where we let the terms with tilde
represent the medium modified quantities.
Here we have defined \cite{sw86,bm90}
\[
   \tilde{m}=m+U_S,
\]
and
\[
      \tilde{E}_{\alpha}=
      \tilde{E}(p_{\alpha})=\sqrt{\tilde{m}^2+{\bf p}_{\alpha}^2}.
\]
The relativistic analog of eq. (\ref{eq:spnrel}) is \cite{bm90}
\begin{equation}
   \tilde{\varepsilon}_{\alpha} =\tilde{E}_{\alpha} +U_V,
   \label{eq:sprelen}
\end{equation}
and the sp potential is given as
\begin{equation}
   u_{\alpha} =\sum_{h\leq k_F} \frac{\tilde{m}^2}{\tilde{E}_h
	       \tilde{E}_{\alpha}}
	\bra{\alpha h}\tilde{G}(\tilde{E}=\tilde{\varepsilon}_{\alpha}
	+\tilde{\varepsilon}_h)\ket{\alpha h}_{\mathrm{AS}},
	\label{eq:urel}
\end{equation}
or, if we wish to express it in terms of the constants $U_S$ and
$U_V$,
we have
\begin{equation}
   u_{\alpha} = \frac{\tilde{m}}{\tilde{E}_{\alpha}}U_S +U_V.
   \label{eq:sppotrel}
\end{equation}
Eq.\ (\ref{eq:spen}) becomes
\begin{equation}
  \begin{array}{ccc}
   \tilde{\varepsilon}_{\alpha}=&\tilde{E}_{\alpha} +U_V,
   &|{\bf p}_{\alpha}|\leq k_M,\\
   &&\\
  =&\tilde{E}_{\alpha},& |{\bf p}_{\alpha}|> k_M ,\\
  \end{array}
\end{equation}
where $k_M=k_F$ gives the traditional BHF sp spectrum while $k_M=ak_F$
with $a>1$ results in the MBHF approach.
In eq. (\ref{eq:urel}), we have introduced the relativistic
$\tilde{G}$-matrix,
which in a partial wave representation is given by
\begin{equation}
   \tilde{G}_{ll'}^{\alpha}(kk'K\tilde{E})=\tilde{V}_{ll'}^{\alpha}(kk')
   +\sum_{l''}\int \frac{d^3 q}{(2\pi )^3}\tilde{V}_{ll''}^{\alpha}(kq)
   \frac{\tilde{m}^2}{\tilde{E}_{\frac{1}{2}K+q}^2}
   \frac{Q(q,K)}{\tilde{E}-2\tilde{E}_{\frac{1}{2}K+q}}
   \tilde{G}_{l''l'}^{\alpha}(qk'K\tilde{E}),
   \label{eq:grel}
\end{equation}
where the relativistic starting energy is
$\tilde{E}=2\tilde{E}_{\frac{1}{2}K+k}$.

Equations (\ref{eq:sprelen})-(\ref{eq:grel}) are solved self-consistently
in the same fashion as in the non-relativistic case, starting
with adequate values for the scalar and vector components
$U_S$ and $U_V$. This iterative scheme is continued until these
parameters show little variation. The calculations are carried out in the
neutron matter rest frame, avoiding thereby a cumbersome
transformation between the two-nucleon center-of-mass system
and the neutron matter rest frame. The additional factors
$\tilde{m}/\tilde{E}$ in the above equations
arise due to the normalization of the
neutron matter spinors $\tilde{w}$, i.e.
$\tilde{w}^{\dagger}\tilde{w}=1$ \cite{bm90}.

Finally, the relativistic version of eq. (\ref{eq:enrel}) reads
\begin{equation}
   {\cal E}/A =
   \frac{1}{A}\left\{ \sum_{h\leq k_F}
   \frac{\tilde{m}m+{\bf p}^2}{\tilde{E}_h}+
   \frac{1}{2}\sum_{hh' \leq k_F}
   \frac{\tilde{m}^2}{\tilde{E}_h\tilde{E}_{h'}}
   \bra{hh'}\tilde{G}
   (\tilde{E}=
    \tilde{\varepsilon}_h +\tilde{\varepsilon}_{h'})
    \ket{hh'}_{\mathrm{AS}}\right\}
    -m.
   \label{eq:erel}
\end{equation}

\subsection{Neutron star equations}
We end this section by presenting the formalism needed in order
to calculate the mass, radius, moment of inertia and gravitational
redshift.
We will assume that the neutron stars we study exhibit an isotropic
mass distribution. Hence, from the general theory of relativity,
the structure of a neutron star is determined through the
Tolman-Oppenheimer-Volkov eqs., i.e.,
\begin{equation}
   \frac{dP}{dr}=
		 - \frac{\left\{\rho (r)+P(r) \right\}
		  \left\{M(r)+4\pi r^3 P(r)\right\}}{r^2- 2rM(r)},
   \label{eq:tov}
\end{equation}
and
\begin{equation}
   \frac{dM}{dr}=
		 4\pi r^2 \rho (r),
   \label{eq:derM}
\end{equation}
where $P(r)$ is the pressure, $M(r)$ is
the gravitional mass inside a radius $r$, and $\rho (r)$ is the
mass energy density. The latter equation is conventionally written
as an integral equation
\begin{equation}
   M(r)=
	4\pi \int_{0}^{r} \rho (r')r'^{2} dr' .
   \label{eq:M}
\end{equation}
In addition, the  main ingredient in a calculation of astrophysical
observables is the equation of state (EOS)
\begin{equation}
   P(n)=
     n^2 \left(\frac{\partial \epsilon}{\partial n}\right),
  \label{eq:P}
\end{equation}
where $\epsilon ={\cal E}/A$ is
the energy per particle and $n$ is the particle
density.
Eqs.\ (\ref{eq:tov}), (\ref{eq:M}) and (\ref{eq:P}) are the basic
ingredients in our calculations of neutron star properties.

The moment of inertia $I$ for a slowly rotating symmetric neutron star
is related to the angular momentum $J$
and the angular velocity $\Omega$ in an
inertial system at infinity through
\begin{equation}
  I=
    \left(\frac{\partial J}{\partial \Omega}\right)_{\Omega =0}=
    \frac{J}{\Omega}.
    \label{eq:I}
\end{equation}
The metric outside a slowly rotating star is taken to be
the Schwarzschild metric
with an additional cross term
\[
   -2\omega r^2 sin^2 \theta d\phi dt,
\]
where $\omega (r)$ is the angular velocity of the local
non-rotating system as measured by an observer in a far-away
inertial system. In order to obtain the moment of inertia $I$, we need
to calculate $J$ and $\Omega$ in eq.\ (\ref{eq:I}). To obtain these
quantities, we introduce the quantity
\[
   f(r) =
	 1-\frac{\omega (r)}{\Omega},
\]
we obtain
\begin{equation}
   \frac{d}{dr}\left(r^4 j(r)\frac{df}{dr}\right)
   +4r^3 \frac{dj}{dr}f(r)=0,
   \label{eq:derf}
\end{equation}
where we have defined
\[
  j(r)=
       \sqrt{1-2M/r}\exp{(-\nu (r)/2)},
\]
and the metric coefficient $\exp{(-\nu (r))}=g_{00}$.

Integrating eq.\ (\ref{eq:derf}) from $r=0$ to $r=R$, assuming
a spherical field in the vacuum outside the star and constraining
the system to the boundary
conditions ($\tilde{\omega}=\Omega -\omega$), i.e.,
\[
  \tilde{\omega}(0)=const, \hspace{1cm}
      \left(\frac{d\tilde{\omega} }{dr}\right)_{r=0}=0,
\]
we get
\begin{equation}
   \tilde{\omega}(r)=\Omega -\frac{2J}{r^3},\hspace{1cm} r > R,
\end{equation}
and
\begin{equation}
   f(r)=1 -\frac{2J}{r^3 \Omega},\hspace{1cm} r > R,
\end{equation}
leading to
\begin{equation}
   \tilde{\omega}(\infty)=\Omega, \hspace{1cm} f(\infty )=1,
\end{equation}
and
\begin{equation}
   J=\frac{1}{6}R^{4}\left(\frac{d\tilde{\omega}}{dr}\right)_{r=R},
\end{equation}
where
$\tilde{\omega}$ is the angular velocity relative to particles with zero
angular momentum.

The moment of inertia $I$ follows from eq. (\ref{eq:I}). To get the
angular momentum $\Omega$, we define the quantity
\begin{equation}
   u=r^4 \frac{d\tilde{\omega}}{dr},
\end{equation}
and obtain
\begin{equation}
   \frac{du}{dr}+\frac{d(lnj)}{dr}\left(u+4r^3 \tilde{\omega}\right)=0,
\end{equation}
and
\begin{equation}
   \frac{d\tilde{\omega}}{dr} =\frac{u}{r^4}.
\end{equation}

For a given equation of state, we get
\begin{equation}
   \frac{d\nu}{dr}=-2\left(\rho +P\right)^{-1}\frac{dP}{dr},
\end{equation}
and $\Omega$ is calculated from $u$ and $\tilde{\omega}$ with the
result
\begin{equation}
   \Omega = \tilde{\omega}(R)+\frac{u(R)}{3R^3}.
\end{equation}

Finally, the gravitational redshift $Z_s$ is given by
\begin{equation}
Z_s=\left(1-\frac{2M(R)}{R}\right)^{-1/2}-1.
\end{equation}

To calculate the total mass, radius, moment of inertia and gravitational
redshift, we employ the EOS defined in eq. (\ref{eq:P}) with the
boundary conditions
\[
  P_c =P(n_c ), \hspace{1cm} M(0)=0,
\]
where we let the subscript $c$ refer to the center of the star, and
$n_c$ is the central density which
is our input parameter in the  calculations
of neutron star properties.

\section{Results}

\subsection{The equation of state}

As mentioned in the introduction,
the replacement of the non-relativistic Schr\"{o}dinger equation by the
Dirac equation offers a quantitative reproduction of the saturation
properties of nuclear matter \cite{bm90}. Central to these results is the
use of modern meson-exchange potentials with a weak tensor force, where the
strength of the tensor force is reflected in the $D$-state probability
of the deuteron. The main differences in the strength of the tensor force
in nuclear matter arises in the
$T=0$ $^3S_1$-$^3D_1$ channel, though other partial waves also give rise
to tensor force contributions.
To derive the equation of state, we  start from the Bonn NN potential
models as they are defined by the parameters of table A.2 in ref.
\cite{mac89}. These potentials are recognized by the labels A, B and
C, with the former carrying the weakest tensor force.
In neutron matter ($T=1$), however,
the important $^3S_1$-$^3D_1$ channel
does not contribute to the energy per particle, and the difference between
the various potentials is expected to be small. This is indeed the case,
as reported by Li {\em et al.} \cite{lmb92} in a recent neutron
matter calculation. We show the same result in the upper
part of fig.\ \ref{fig:fig1}.
There we plot the BHF results using the non-relativistic
Schr\"{o}dinger equation to describe the sp motion, see the discussion
in subsec.\ 2.1. On the scale of fig.\ \ref{fig:fig1}, there is hardly
any difference. Further, with the existing uncertainty of neutron
star properties, the rather small
differences between the energies derived
from the  Bonn A, B and C potentials are not expected to be
significant, see also the discussion below. A similar conclusion
is reached when we perform the relativistic model-space BHF calculations.
Thus, in our presentation of neutron star properties, we will use
the EOS derived from the Bonn A potential, mainly because this is the
potential which gives the
best reproduction of the saturation properties of
nuclear matter.
\begin{figure}[hbtp]
\vspace{14cm}
\caption{Energy per neutron $E/A$ as function of density
$n$ for various many-body
approaches. The upper part of the figure shows  results obtained
with the non-relativistic BHF approach for the Bonn A (solid line),
B (dashed line) and C (dotted line)
potentials. In the lower part  non-relativistic MBHF
(solid line),  non-relativistic BHF (dashed line) and  relativistic
MBHF (dotted line) results  are shown.
All the last results have been obtained
with the Bonn A potential.}
\label{fig:fig1}
\end{figure}
In the lower part of fig.\ \ref{fig:fig1} we compare first the
non-relativistic BHF and MBHF results. As can be seen from the
corresponding curves which are a solid line and
a dashed line for the non-relativistic
MBHF and BHF results, respectively, there is hardly any difference.
This is again primarily due to the fact that we have no contributions
from the $^3S_1$-$^3D_1$ channel in neutron matter. Actually, nuclear
matter calculations with the MBHF method \cite{km83,lhko92} show
that the most significant difference between the MBHF and BHF
approaches arises in the $^3S_1$-$^3D_1$ channel, with differences
of the order of $ 2$ MeV. This can be understood from
the fact that the MBHF approach with $k_M > k_F$
introduces intermediate state contributions of shorter range,
contributions which are predominantly accounted for by the
$^3S_1$-$^3D_1$ channel. The fact that the MBHF and the BHF
approaches give similar results for neutron matter is a gratifying
property, though the MBHF method\footnote{The equations of state
we  will employ in the derivation of neutron star
properties have all been obtained with the MBHF approach,
both the non-relativistic and the relativistic EOS.}
is to be preferred since it allows
for a consistent treatment of higher-order contributions
in the perturbative expansion, such as the summation of the
particle-particle-hole-hole
 ring diagrams without double-counting problems. The investigations
of such effects are the scope of a future work. Here we limit our
attention to first order in the interaction $G$. However,
the most important contribution from ring diagrams stems from the
$^3S_1$-$^3D_1$ channel, which is
absent in our neutron matter model. Thus,
the most significant differences
between various equations of state, are
due to relativistic effects.
This is clearly
seen from the lower part of fig.\ \ref{fig:fig1}. The
relativistic MBHF calculation gives an increased
repulsion at higher densities
and correspondingly stiffer EOS than the non-relativistic
approaches. In this figure we plot only the relativistic MBHF results,
since they are similar to the relativistic BHF results\footnote{Our
relativistic results are also similar to those reported by Li
{\em et al.} \cite{lmb92}.}, in analogy to the non-relativistic
results discussed above. Using the non-relativistic and the relativistic
equations of state, we wish to study how these two extremes reproduce
various neutron star properties. Note also that within the
Dirac-Brueckner
approach, the Bonn A potential reproduces the empirical nuclear matter
binding energy and saturation density \cite{bm90}. This gives a more
consistent approach to pure neutron  matter. The reader should however
keep in mind that there are several mechanisms (to be discussed in
section 4) which may reduce the stiffness of the above equations of state.

The equation of state can then easily be obtained through the use of eq.\
(\ref{eq:P}). The pressure can also be fitted numerically by the
following polynomials;
$$
    P(n)=0.048376\times n^{4/3} -0.036764\times n^{5/3}
		+4.959151\times n^2
$$
\begin{equation}
	 	-21.095163\times n^{7/3} +32.922084\times n^{8/3}
                -14.213806\times n^3,
\end{equation}
for the non-relativistic EOS with the Bonn A potential
obtained with the MBHF approach, and
$$
    P(n)= -1271.125\times n^{4/3} +2010.914\times n^{5/3}
		+29279.687\times n^2
$$
$$
           -59189.0540\times n^{7/3}
                      -403313.402\times n^{8/3}
                     +2067381.296\times n^3
$$
$$
       -4162678.316\times n^{10/3}
          +4400107.684\times n^{11/3}
$$
\begin{equation}
         -2422293.924\times n^4  +550123.976 \times n^{13/3},
\end{equation}
for the relativistic EOS
obtained with the MBHF approach. The pressure is given in units
of [$10^{34}$ N/m$^2$]. The range of validity for these two
equations of state is $0.1$ fm$^{-3} \leq n \leq 0.8 $fm$^{-3}$.

\subsection{Mass, radius, moment of  inertia and  surface gravitational
redshift}

To calculate mass, radius, moment of  inertia and  surface gravitational
redshift we need the EOS for all relevant densities. The equations
of state derived in the previous subsection have a limited range,
$0.1$ fm$^{-3} \leq n \leq 0.8 $fm$^{-3}$. We
must therefore include equations of state for other densities as well.
These equations of state are discussed below.

For the lowest densities, we use the
 Haensel-Zdunik-Dobaczewski (HZD) \cite{HZD}
equation of state. This equation of state
 is obtained in the following way: The pressure is fitted
by a polynomial consisting of 9 terms, i.e.,
\begin{equation}
P(X)=\sum_{i=1}^9C_iX^{l_i},
\end{equation}
\noindent
where
\begin{equation}
X=1.6749\times10^5n,
\end{equation}
\par\noindent
$n$ is given in [fm$^{-3}$], and the values
\[
n= 0.077,~~ 0.154,~~ 0.395,~~ 0.762,~~ 1.575,
3.147,~~ 6.443,~~ 12.240,~~ 26.551,
\]
in [$10^{-5}$/fm$^3$], are chosen to give the coefficients $C_i$.
The corresponding equations are solved by matrix inversion, and we obtain
$$
P(X)=8.471521942X^{1/3}-40.437728191X^{2/3}+74.927783479X
$$
$$
-67.102601796X^{4/3}+30.011422630X^{5/3}-4.207322319X^{2}
$$
\begin{equation}
-1.419954871X^{7/3}+0.589441363X^{8/3}-0.060468689X^{3},
\end{equation}
where $P(X)$ is given in units of
[$10^{27}$ N/m$^2$] and in the density range of
$2\times10^{-6}$ fm$^{-3}<n< 2.84\times 10^{-4}$ fm$^{-3}$.
We need all the decimals in the different terms to get an accuracy of
at least two decimals in the net equation.

The  Baym-Bethe-Pethick (BBP) \cite{BBP}
equations of state are taken from
{\O}verg{\aa}rd and {\O}stgaard \cite{OO}.
The given data are fitted by two
five-term polynomials to give (BBP-1)
$$
P(n)=4.3591n^{4/3}-122.4841n^{5/3}+1315.2746n^{2}
$$
\begin{equation}
-6180.0702n^{7/3}+10659.0049n^{8/3},   \label{eq:9}
\end{equation}
where $P(n)$ is given  in units of [$10^{34}$ N/m$^2$] for $n$
 in the density range of
$0.00027$ fm$^{-3} <n <0.0089$ fm$^{-3}$,
and ( BBP-2)
$$
P(n)=0.092718n^{4/3}-0.035382n^{5/3}+1.193525n^{2}
$$
\begin{equation}
-2.424555n^{7/3}+2.472867n^{8/3},
\end{equation}
where $P(n)$ is given  in units of [$10^{34}$ N/m$^2$] for $n$
in the density range of
$0.0089$ fm$^{-3} <n <0.3$ fm$^{-3}$.

The Arntsen- {\O}verg{\aa}rd
(A{\O}-5) \cite{OO}
equation of state is given by a five-term polynomial, i.e.,
$$
P(n)= 9.4433n^{5/3}-34.6909n^{2}+102.6575n^{8/3}
$$
\begin{equation}
-87.6158n^{3}+14.3549n^{11/3},
\end{equation}
where $P(n)$ is given  in units of [$10^{34}$ N/m$^2$] for $n$
in the density range of
$0.4$ fm$^{-3} <n <3.6$ fm$^{-3}$.

The Pandharipande - Smith (PS) \cite{PS} equation of state is taken from
{\O}verg{\aa}rd and {\O}stgaard \cite{OO}. The given data are fitted by a
five-term polynomial to give
$$
P(n)=4.0378n^{4/3}-27.853n^{5/3}+52.0859n^{2}
$$
\begin{equation}
-20.7073n^{7/3}+5.5808n^{8/3},
\end{equation}
where $P(n)$ is given  in units of [$10^{34}$ N/m$^2$] for $n$
in the density range of
$0.1$ fm$^{-3} <n <3.6$ fm$^{-3}$.

For our non-relativistic  equation
of state (BEHO{\O}-nr)  we find that the
following equations of state are the best to cover the whole
range of  densities in a neutron star, and  we use:

\noindent
HZD in the density range of
\[
 n< 0.000256,
\]
\noindent
BBP-1 in the density range of
\[
0.000256\le n < 0.003892,
\]
\noindent
BBP-2 in the density range of
\[
 0.003892\le n<0.08,
\]
BEHO{\O}-nr in the density range of
\[
0.08 \le n<0.8,
\]
\noindent
A{\O}-5 in the density range of
\[
 0.8\le n<3.46,
\]
\noindent
and PS in the density range of
\[
n  \ge 3.46,
\]
where $n$  is given in  units of [fm$^{-3}$].

For our relativistic equation of state (BEHO{\O}-r), we have coupled
the following equations of state:

\noindent
HZD in the density range of
\[
 n< 0.000256,
\]
\noindent
BBP-1 in the density range of
\[
0.000256 \le n< 0.003892,
\]
\noindent
BBP-2 in the density range of
\[
 0.003892 \le n<0.115,
\]
BEHO{\O}-gr in the density range of
\[
0.115\le n<0.7,
\]
\noindent
and PS in the density range of
\[
n  \ge 0.7,
\]
where $n$  is given in units of [fm$^{-3}$].
These equations are chosen among 12 published equations of state,
and they seem to be  the best ones coupled together
in the total density range.
Total masses,  radii, moments of inertia and surface gravitational
redshifts are then calculated, and  parameterized as  functions of
 the central density
$n_c$.  Table~1 gives  results related to our non-relativistic
equation of state and  table~2 gives  results for  the relativistic case.
In both  tables column 1 refers to
the central density  of the neutron star,
column 2 to  the total mass, column 3 to the radius,
column 4 to the moment of inertia,  and column 5 refers to
the  gravitational
redshift. Fig.~2  shows
 the total mass  and fig.~3 the radius versus the   central density.
  Fig.~4 shows the mass versus the radius.
Fig.~5 shows  the moment of inertia
and fig.~6 the gravitational redshift
versus the  total mass of the  star.

\section{Discussions and conclusions}

{}From table~1, table~2, and figs.~2--4 we find a  maximum mass of
\[
M_{\mathrm{max}}\approx 1.46 M_{\odot},
\]
\noindent at a central density of $n_c
\approx 2.15$ fm$^{-3}$  with a radius
$ R\approx 7.45 $ km for our
non-relativistic equation of state.  The maximum mass for our
relativistic equation of state is
\[
M_{\mathrm{max}}\approx 2.37 M_{\odot},
\]
\noindent at a central density of $n_c
\approx 0.8 $ fm$^{-3}$ with a radius
$R \approx 12.12$ km.
\begin{table}
\caption{Neutron star observables as functions of the central density
$n_c$ obtained with
the non-relativistic equation of state
(BEHO{\O}-nr). $M$ is the total mass, $R$ the radius, $I$ the moment
of inertia and $Z_s$ the gravitational redshift.}
\begin{center}
\begin{tabular}{ccccc}
&&&&\\ \hline
&&&&\\
 $n_c$ [fm$^{-3}$] &$M[M_{\odot}$] &$R$[km]&$I[10^{38}$kg m$^2$]&$Z_s$ \\
&&&&\\ \hline
&&&&\\
        0.500      & 0.580     &10.18&      0.311  &0.011    \\
        0.650      & 0.789     &9.877&      0.453  &0.144    \\
        0.800      & 0.957     &9.641&      0.567  &0.189    \\
        0.950      & 1.113     &9.332&      0.663  &0.243    \\
        1.100      & 1.230     &9.057&      0.730  &0.292    \\
        1.250      & 1.312     &8.807&      0.769  &0.336    \\
        1.400      & 1.369     &8.570&      0.787  &0.376    \\
        1.550      & 1.408     &8.339&      0.790  &0.412    \\
        1.700      & 1.434     &8.111&      0.781  &0.447    \\
        1.850      & 1.449     &7.887&      0.763  &0.479    \\
        2.000      & 1.457     &7.665&      0.739  &0.510    \\
        2.150      & 1.458     &7.450&      0.711  &0.539    \\
        2.300      & 1.453     &7.241&      0.681  &0.567    \\
        2.450      & 1.444     &7.o41&      0.649  &0.592    \\
        2.600      & 1.431     &6.885&      0.617  &0.615    \\
        2.750      & 1.416     &6.681&      0.586  &0.635    \\
        2.900      & 1.399     &6.521&      0.556  &0.652    \\
        3.050      & 1.381     &6.377&      0.528  &0.666    \\
        3.200      & 1.363     &6.247&      0.502  &0.677    \\
        3.350      & 1.344     &6.132&      0.477  &0.684    \\
        3.425      & 1.335     &6.077&      0.466  &0.687    \\
        &&&&\\ \hline
\end{tabular}
\end{center}
\end{table}
\begin{table}
\caption{Neutron star observables as functions of the central density
$n_c$ obtained with
the relativistic equation of state
(BEHO{\O}-r). $M$ is the total mass, $R$ the radius, $I$ the moment
of inertia and $Z_s$ the gravitational redshift.}
\begin{center}
\begin{tabular}{ccccc}
&&&&\\ \hline
&&&&\\
$n_c$ [fm$^{-3}$]  &$M[M_{\odot}$]&$R$ [km]&$I [10^{38}$kg m$^2$]&$Z_s$ \\
&&&&\\ \hline
&&&&\\
        0.500      & 2.227     &13.41     & 3.431&0.401      \\
        0.650      & 2.356     &12.74     & 3.412&0.485      \\
        0.800      & 2.370     &12.14     & 3.169&0.538      \\
        0.950      & 2.337     &11.60     & 2.880&0.572      \\
        1.100      & 2.286     &11.16     & 2.607&0.591      \\
        1.175      & 2.258     &10.97     & 2.482&0.598      \\
        1.250      & 2.229     &10.79     & 2.367&0.602      \\
        1.400      & 2.173     &10.49     & 2.161&0.606      \\
        1.550      & 2.120     &10.23     & 1.986&0.605      \\
        1.700      & 2.070     &10.01     & 1.838&0.603      \\
        1.850      & 2.025     &9.833     & 1.713&0.596      \\
        2.000      & 1.983     &9.679     & 1.606&0.591      \\
        2.150      & 1.946     &9.551     & 1.515&0.584      \\
        2.300      & 1.912     &9.442     & 1.437&0.577      \\
        2.450      & 1.881     &9.351     & 1.370&0.570      \\
        2.600      & 1.853     &9.272     & 1.312&0.562      \\
        2.750      & 1.827     &9.208     & 1.261&0.555      \\
        2.900      & 1.805     &9.155     & 1.218&0.547      \\
        3.050      & 1.784     &9.111     & 1.180&0.540      \\
        3.200      & 1.766     &9.076     & 1.147&0.533      \\
        3.350      & 1.749     &9.045     & 1.118&0.527      \\
        3.425      & 1.741     &9.033     & 1.105&0.524      \\
        &&&&\\ \hline
\end{tabular}
\end{center}
\end{table}
These results agree very well with ``experimental results" from observations
of binary pulsars, which give neutron star masses of \cite{OO,PAL}
\[
1.0M_{\odot} < M_{\mathrm{max}} < 2.2 M_{\odot},
\]
\noindent or possibly \cite{glen88,TW,JR}
\[
1.4M_{\odot} < M_{\mathrm{max}} < 1.85 M_{\odot},
\]
\noindent
and imply that stars calculated with stiff
equations of state have greater maximum mass, lower central density
and thicker crust than stars obtained with soft equations of state.
\begin{figure}
\vspace{8cm}
\caption{Total mass as function of  central density
for neutron stars. NR indicates the non-relativistic equation of state
(BEHO{\O}-nr) and R the relativistic equation of state
(BEHO{\O}-r).}\label{fig:fig2}
\end{figure}
\begin{figure}
\vspace{8cm}
\caption{Total radius as function of  central density
for neutron stars. NR indicates the non-relativistic equation of state
(BEHO{\O}-nr) and R the relativistic equation of state
(BEHO{\O}-r).}\label{fig:3}
\end{figure}
\begin{figure}
\vspace{8cm}
\caption{Total mass as function of  radius
for neutron stars. NR indicates the non-relativistic equation of state
(BEHO{\O}-nr) and R the relativistic equation of state
(BEHO{\O}-r).}\label{fig:fig4}
\end{figure}
\begin{figure}
\vspace{8cm}
\caption{ Moment of inertia as function of  total mass
for neutron stars. NR indicates the non-relativistic equation of state
(BEHO{\O}-nr) and R the relativistic equation of state
(BEHO{\O}-r).}\label{fig:fig5}
\end{figure}
\begin{figure}
\vspace{8cm}
\caption{Surface gravitational redshift as function of  total mass
for neutron stars. NR indicates the non-relativistic equation of state
(BEHO{\O}-nr) and R the relativistic equation of state
(BEHO{\O}-r).}\label{fig:fig6}
\end{figure}

At present, no reliable measurements of the  radius of a neutron star exist.
But general estimates give \cite{OO}
\[
R \approx 9 \mathrm{km}.
\]
\noindent If this estimate is close to the true value, then the results
from our non-relativistic equation of state may look more
reasonable than those from the  relativistic one. However,
 theoretical calculations of the radius of neutron stars
can not be confirmed very well   by observational data, and  are more dependent
than the total mass on the  low-density equation of state.

{}From table~1, table~2, and fig.~5 we see that
our models give a value of
\[
I(M_{\mathrm{max}})= 0.71 \times 10^{38} \mathrm{kg m^2},
\]
\noindent
for the moment of inertia of a neutron star of maximum mass
\noindent
and
for the non-relativistic equation of state, and
\[
I(M_{\mathrm{max}})= 3.17 \times 10^{38}\mathrm{ kg m^2},
\]
\noindent for the  relativistic equation of state, while
\[
I_{\mathrm{max}}=0.80 \times 10^{38} \mathrm{kg m^2},
\]
\noindent
and
\[
I_{\mathrm{max}}=3.47 \times 10^{38} \mathrm{kg m^2},
\]
\noindent
for our non-relativistic and  relativistic equations of state, respectively.
These values  are not contradictory to  observations, and are
 consistent with
the expansion of the Crab nebula and the luminosity  and the loss of
rotational energy from the Crab pulsar \cite{MO}.

{}From fig.~6 we see that the gravitational surface redshift is not
strongly affected by the different equations of state. This is because
the density profiles of the stars are such that their
surface gravities are almost the same. A measurement of the redshift
can therefore not be used to distinguish between different types of stars or
equations of state. It is, however, possible that the slowing down of pulsars
and the corresponding glitches can give some
information about the internal structure.

For some other compact astrophysical objects like, for  example, white dwarfs
or
supermassive black holes in  galactic nuclei,
theoretical calculations of mass, radius, etc. can provide
constraints on physical models. Theoretical calculations for neutron stars,
however,
can not really give such  information due to uncertainties
of their internal structure. Instead, we may obtain
some information about the nuclear physics of the  interior of the star.

Data on the nuclear equation of state can, in principle, be obtained
from several different sources such as the monopole resonance in nuclei,
 high energy nuclear collisions, supernovae, and neutron stars.
Until recently it has, for instance, been assumed that the compression modulus
was reasonably well known from the analysis of the giant monopole resonance
in nuclei \cite{BGG,B,TKBM}. Later, however, these results were
questioned by the authors of ref.\ \cite{glen88,BO}.

Supernova simulations seem to require an equation of state which is too
soft to support some observed masses of neutron stars, if sufficient
 energy shall be released to make the ejection mechanism work
\cite{BCK,WW,BBBCK}.
Supernova explosions can then probably not give a reliable
constraint on the nuclear equation of state. Some analyses of high
 energy nuclear collisions, however, have indicated a moderately stiff
or very stiff equation of state \cite{SGWK,SG,MHS}, although
ambiguities have been observed \cite{GBG,ARPSG}.
Various nuclear data and neutron star masses then seem to favour a rather high
 compression modulus of $K\sim 300$ MeV \cite{glen88,SBBOWH,SSGB}.
 No definite statements can be made, however, about the equation of state at
high densities, except that  the neutron star equation of
state  should probably be moderately stiff to support neutron star masses
 up to approximately 1.85 $M_\odot$ \cite{JR}.

With the above observations, one may be tempted to state that our
relativistic EOS is too stiff, since the predicted mass
$M_{max}\approx 2.37M_\odot$
and radius are larger than the estimated values. However, there are several
mechanisms which serve to soften a given EOS. Amongst these,
pions may be likely to condense in neutron star matter, favoured by
charge neutrality because neutrons at the top of the  Fermi sea could decay
to protons plus electrons. Condensation then could possibly occur
if the pion energy becomes degenerate with the normal state, but the
real situation is still not clear \cite{MO88}. Also,
kaon condensation through the
s-wave interaction of kaons with nucleons may be energetically
favourable in the interior of neutron stars \cite{TPL,BLRT}. Kaon
condensation would, like pion condensation, increase the proton
abundance of matter \cite{MO88},
 and could cause a rapid cooling of neutron stars via
the direct URCA process. Kaon condensed neutron star cores  may also
undergo a phase transition to  strange quark matter, and neutrino trapping
in newly formed neutron star matter could shift both the kaon
condensation and the quark matter transition to higher densities.
The equation of state for neutron stars would be softened considerably due to
pion or
kaon condensation because of the lower symmetry energy of nuclear matter,
and maximum masses are then reduced correspondingly from the cases  with no
condensates. This should have  important implications for the
formation of black holes in stellar collapse, and the number of black holes
in the Galaxy
should then be substantially higher than estimated earlier.

Further  processes
 which can soften the  equation of state
are also conversion of nucleons to  hyperons or a phase transition
to quark matter at high densities, which  would both   lower the energy
by an increase in the number of degrees of freedom.
However, for a neutron star to resist  the centrifugal forces from very
fast rotation, the equation of state should be soft at low and
intermediate densities  and stiff at high densities, which would not
fit very well with the concept of  quark matter in
 hybrid stars \cite{wgw91}.

In summary, in this work we have calculated the EOS for neutron
matter using the recent meson-exchange potential models of the Bonn
group. Both a non-relativistic and a relativistic Brueckner-Hartree-Fock
procedure were employed in order to derive the equation of state,
which is the basic input quantity
in  neutron star calculations since it connects
the nuclear physics and the astrophysics. Of importance here is the fact
that a relativistic nuclear matter calculation with the Bonn A potential
\cite{bm90} meets the empirical nuclear matter data, a feature not accounted
for by non-relativistic calculations.
The relativistic effects become important at densities around and higher
than the saturation density for nuclear matter, and their main effect is
to stiffen the EOS at these densities. This mechanism is due to the fact that
the relativistic effective mass of the nucleon becomes smaller compared
to the free mass, an effect which in turn enhances the repulsive spin-orbit
term.

Albeit
the description of the matter
inside the star is
a complicated many-body problem,
we have in this work used equations of state for neutron matter only,
in order to assess the importance of relativistic effects in neutron
star calculatons. Our conclusions are that the relativistic EOS yields
too stiff an EOS, however, many-body effects not included here, may
soften the EOS and bring the relativistic results close to the empirical
values
for mass and radius. The other observables like moment of inertia and
gravitational redshifts are in good agreement with the accepted values.

\bigskip

Many discussions with Ruprecht Machleidt on the Dirac-Brueckner
approach are greatly acknowledged.

\clearpage



\begin{thebibliography}{99}
\bibitem{chris92a} C.J.\ Pethick and D.G.\ Ravenhall, Phil.\ Trans.\
Roy.\ Soc.\ Lond.\ {\bf A341} (1992) 17
\bibitem{wg91} F.\ Weber and N.K.\ Glendenning,
{\em Hadronic matter and rotating relativistic neutron stars},
(World Scientific, Singapore, in press)
\bibitem{chris92b} C.J.\ Pethick, Rev.\ Mod.\ Phys.\ {\bf 64} (1992)1133
\bibitem{mac89} R.\ Machleidt, Adv.\ Nucl.\ Phys.\ {\bf 19} (1989) 189
\bibitem{mac93} R.\ Machleidt and G.Q.\ Li, to appear in Physics Reports
\bibitem{sw86} S.D.\ Serot and J.D.\ Walecka,
Adv.\ Nucl.\ Phys.\ {\bf 16}  (1986) 1
\bibitem{wff88} R.B.\ Wiringa, V.\ Fiks and A.\ Fabrocini, Phys.\ Rev.\
{\bf C38} (1988) 1010
\bibitem{dm92} W.H.\ Dickhoff and H.\ M\"{u}ther, Rep.\ Prog.\ Phys.\
{\bf 55} (1992) 1947
\bibitem{km83} T.T.S.\ Kuo  and Z.Y.\ Ma,
Phys.\ Lett.\ {\bf B127} (1983) 123;
T.T.S.\ Kuo  and Z.Y.\ Ma, in
{\em Nucleon-Nucleon Interaction and Nuclear
Many-body Problems}  (edited by S.S.\ Wu and T.T.S.\ Kuo)
(World Scientific:
Singapore) p. 178;
T.T.S.\ Kuo, Z.Y.\ Ma and R. Vinh Mau,
Phys.\ Rev.\ {\bf C33} (1986) 717
\bibitem{ray91} L.\ Ray, G.W.\ Hoffman and W.R.\ Cooker, Phys. Rep. {\bf 212}
(1991) 223
\bibitem{hnm92} T.\ Nikolaus, T.\ Hoch and D.G.\ Madland, Phys.\ Rev.\
{\bf C46} (1992) 1757
\bibitem{br78} R.\ Brockmann, Phys.\ Rev.\ {\bf C18} (1978) 1510
\bibitem{bm90} R.\ Brockmann and R.\ Machleidt,
Phys.\ Rev.\ {\bf C42} (1990) 1965
\bibitem{mmb90} H.\ M\"{u}ther, R.\ Machleidt and R.\ Brockmann, Phys.\ Rev.\
{\bf C42} (1990) 1981
\bibitem{lmb92} G.Q.\ Li, R.\ Machleidt and R.\ Brockmann,
Phys.\ Rev.\ {\bf C46} (1992) 2782
\bibitem{glen88} N.K.\ Glendenning, Phys.\ Rev.\ {\bf C37} (1988) 2733
\bibitem{wgw91} F.\ Weber, N.K.\ Glendenning and M.K.\ Weigel,
Astrophys. Journ. {\bf 373} (1991) 579
\bibitem{ms89} C.\ Mahaux  and R.\ Sartor,
Phys.\ Rev.\ {\bf C 40} (1989) 1833;
C.\ Mahaux, P.E.\ Bortignon, R.A.\ Broglia and C.H.\ Dasso,
 Phys.\ Rep.\ {\bf 120} (1985) 1
\bibitem{shk87} H.Q.\ Song, S.D.\ Yang and T.T.S.\ Kuo,
Nucl.\ Phys.\ {\bf A462} (1987) 491
\bibitem{jkm89} M.F.\ Jiang, R.\ Machleidt and T.T.S.\ Kuo,
Phys.\ Rev.\
{\bf C41} (1989) 2346;
M.F.\ Jiang, T.T.S.\ Kuo  and H.\ M\"{u}ther,
Phys.\ Rev.\ {\bf C40} (1989) 1836
\bibitem{brook} L.S.\ Celenza and C.M.\ Shakin, {\em Relativistic
Nuclear Physics: Theories of Structure and Scattering}, Vol. 2
of Lecture Notes in Physics, (World Scientific, Singapore, 1986)
\bibitem{hm87} B. ter Haar and R. Malfliet, Phys. Rep. {\bf 149} (1987)
207
\bibitem{hs87} C.J.\ Horowitz and B.D.\ Serot, Nucl.\ Phys.\ {\bf A464}
(1987) 613
\bibitem{ser92} B.D.\ Serot, Rep.\ Prog.\ Phys.\ {\bf 55} (1992) 1855
\bibitem{thomp70} R.H.\ Thompson, Phys.\ Rev.\ {\bf D1} (1970) 110
\bibitem{iz80} C.\ Itzykson and J.-B.\ Zuber, {\em Quantum Field theory},
(McGraw-Hill, New York, 1980)
\bibitem{lhko92} L.\ Engvik, M.\ Hjorth-Jensen, T.T.S.\ Kuo and
E.\ Osnes, to be published
\bibitem{HZD} P. Haensel, J.L.\ Zdunik and J.\ Dobaczewski,
Astron.\ Astrophys.\ {\bf 222} (1989) 353
\bibitem{BBP} G.\ Baym, H.A.\ Bethe and C.J.\ Pethick,
 Nucl.\ Phys.\ {\bf A175} (1971) 225
\bibitem{OO} T.\ {\O}verg{\aa}rd and E.\ {\O}stgaard,  Can.\ Journ.\ Phys.\
{\bf 69}
(1991) 8
\bibitem{PS} V.R.\ Pandharipande and R.A.\ Smith, Nucl.\ Phys.\ {\bf A237}
(1975) 507
\bibitem{PAL} M.\ Prakash, T.L.\ Ainsworth and J.M.\ Lattimer, Phys.\ Rev.\
Lett.\
{\bf 61} (1988) 2518
\bibitem{TW} J.H.\ Taylor and J.M.\ Weisberg, Astrophys.\ Journ.\ {\bf 345}
(1989) 434
\bibitem{JR} P.C.\ Joss and S.A.\ Rappaport, Ann.\ Rev.\ Astron.\ Astrophys.\
{\bf 22} (1984) 537
\bibitem{MO} T.\ M{\o}lnvik and E.\ {\O}stgaard,  Nucl.\ Phys.\ {\bf A437}
(1985) 239
\bibitem{BGG} J.P.\ Blaizot, D.\ Gogny and B.\ Grammiticos, Nucl.\ Phys.\
{\bf A265} (1976) 315
\bibitem{B} J.P.\ Blaizot, Phys.\ Rep.\ {\bf 64} (1980)171
\bibitem{TKBM} J.\ Treiner, H.\ Krevine, O.\ Bohigas  and J.\ Martorell,
Nucl. Phys. {\bf A317} (1981) 253
\bibitem{BO} G.E.\ Brown and E.\ Osnes, Phys.\ Lett.\ {\bf B159} (1985) 223
\bibitem{BCK} E.\ Baron, J.\ Cooperstein and S.\ Kahana, Phys.\ Rev.\ Lett.\
{\bf 55}
(1985) 126
\bibitem{WW} S.E.\ Woosley and T.A.\ Weaver, Ann.\ Rev.\ Astron.\ Astropys.\
{\bf 24} (1986) 205
\bibitem{BBBCK} E.\ Baron, H.A.\ Bethe, G.E.\ Brown, J.\ Cooperstein
 and S.\ Kahana, Phys.\ Rev.\ Lett.\ {\bf 59} (1987) 736
\bibitem{SGWK} M.\ Sano, M.\ Gyulassy, M.\ Wakai and Y.\ Kitazoe,
Phys.\ Lett.\ {\bf B156} (1985) 27
\bibitem{SG} H.\ Stocker and W.\ Greiner, Phys.\ Rep.\ {\bf 137} (1986) 277
\bibitem{MHS} J.J.\ Molitoris, D.\ Hahn and H.\ Stocker, Nucl.\ Phys.\ {\bf
A447}
(1985) 13c
\bibitem{GBG} C.\ Gale, G.\ Bertsch and S.\ Das Gupta, Phys.\ Rev.\ {\bf C35}
(1987) 415
\bibitem{ARPSG} J.\ Aichelin, A.\ Rosenhauer, G.\ Peilert, H.\ Stocker and
W. Greiner, Phys.\ Rev.\ Lett.\ {\bf 58} (1987) 1926
\bibitem{SBBOWH} M.M.\ Sharma, W.T.A.\ Borghols, S.\ Brandenburg, S.\ Crona,
A.\
van der Woude and M.N.\ Harakeh, Phys.\ Rev.\ {\bf C38} (1988) 2562
\bibitem{SSGB} M.M.\ Sharma, W.\ Stocker, P.\ Gleissl and M.\ Brack,
Nucl.\ Phys.\ {\bf A504} (1989) 337
\bibitem{WG} F.\ Weber and N.K.\ Glendenning, Astrophys.\ Journ.\ {\bf 390}
(1992) 541
\bibitem{MO88} R.\ Mittet and E. {\O}stgaard, Phys.\ Rev.\ {\bf C37} (1988)
1711
\bibitem{TPL} V.\ Thorsson, M.\ Prakash and J.\ Lattimer,
NORDITA preprint, no.\ {\bf 29} (1993)
\bibitem{BLRT} G.E.\ Brown, C.-H.\ Lee, M.\ Rho and V.\ Thorsson,
NORDITA preprint, no.\ {\bf 30} (1993)
\end{thebibliography}
\end{document}